\documentclass[fleqn,usenatbib]{mnras}
\usepackage{newtxtext,newtxmath}
\usepackage[T1]{fontenc}

\DeclareRobustCommand{\VAN}[3]{#2}
\let\VANthebibliography\thebibliography
\def\thebibliography{\DeclareRobustCommand{\VAN}[3]{##3}\VANthebibliography}

\usepackage[british]{babel}

\usepackage{amsmath}
\usepackage{graphicx}

\title[Main sequence and radial growth]{Implications of a spatially resolved main sequence for the size evolution of star forming galaxies}
\author[G. Pezzulli]{Gabriele Pezzulli$^{1}$\thanks{E-mail:pezzulli@astro.rug.nl}\\
$^{1}$ Kapteyn Astronomical Institute, University of Groningen, Landleven 12, 9747 AD, Groningen, The Netherlands}

\date{Accepted 2021 September 29. Received 2021 September 26; in original form 2021 August 13}
\pubyear{2021}

\begin{document}
\label{firstpage}
\pagerange{\pageref{firstpage}--\pageref{lastpage}}
\maketitle



\begin{abstract}
\noindent
Two currently debated problems in galaxy evolution, the fundamentally local or global nature of the main sequence of star formation and the evolution of the mass-size relation of star forming galaxies (SFGs), are shown to be intimately related to each other. As a preliminary step, a \emph{growth function} $g$ is defined, which quantifies the differential change in half-mass radius per unit increase in stellar mass ($g = d \log R_{1/2}/d \log M_\star$) due to star formation. A general derivation shows that $g = K\Delta(sSFR)/sSFR$, meaning that $g$ is proportional to the \emph{relative} difference in specific star formation rate between the outer and inner half of a galaxy, with $K$ a dimensionless structural factor for which handy expressions are provided. As an application, it is shown that galaxies obeying a fundamentally local main sequence also obey, to a good approximation, $g \simeq \gamma n$, where $\gamma$ is the slope of the \emph{normalized} local main sequence ($sSFR \; \propto \; \Sigma_\star^{-\gamma}$) and $n$ the Sersic index. An exact expression is also provided. Quantitatively, a fundamentally local main sequence is consistent with SFGs growing along a stationary mass-size relation, but inconsistent with the continuation at $z=0$ of evolutionary laws derived at higher $z$. This demonstrates that either the main sequence is not fundamentally local, or the mass-size relation of SFGs has converged to an equilibrium state some finite time in the past, or both. \end{abstract}


\section[Introduction]{Introduction}\label{sec:Introduction}

How closely is the evolution of star forming galaxies (SFGs) dictated by large-scale cosmological processes? To what extent are instead SFGs able to regulate their own growth by means of more local mechanisms? These are common questions underlying several ongoing investigations in galaxy evolution and remind us of the importance to establish connections between various sub-fields. Two problems in particular are often studied separately, but are in fact -- as the present study will try to highlight -- tightly connected: the local or global nature of the main sequence of star formation and the evolution of the mass-size relation of SFGs.

\subsection{Evolution of the mass-size relation}\label{subsec:intro_masssize}

In its simplest and most common form, the mass-size relation connects the stellar mass of a galaxy to its \emph{half-mass} radius $R_{1/2}$, with the latter often approximated by the half-light or \emph{effective} radius $R_\textrm{eff}$, measured in the (rest-frame) optical or, preferably, near infra-red (e.g.\ \citealt{Lange+15}; \citealt{Lelli+16}; \citealt{Wu+18}). If the evolution of galaxies is mainly governed by the growth of their parent haloes, it is then expected that the relation should evolve with cosmic time (e.g.\ \citealt{MMW98}; \citealt{Dutton+11}). This expectation appears, at least qualitatively, confirmed by observational studies based on the measurement of (rest-frame optical) half-light radii as a function of stellar mass and redshift (e.g.\ \citealt{vdW+14}; \citealt{Paulino-Afonso+17}; \citealt{Mowla+19}; \citealt{Nedkova+21}). Theoretical efforts are on their way to clarify whether the observed evolution merely reflects the size evolution of the parent haloes (e.g.\ \citealt{Huang+17}) or other processes play a major role as well (e.g.\ \citealt{Somerville+18} and references therein). Observational constraints on an evolving mass-size relation are also a valuable test for cosmological hydrodynamical simulations (e.g.\ \citealt{Furlong+17}; \citealt{Genel+18}).

Given the importance, for galaxy evolution theory, of a redshift-dependent mass-size relation, considerable scrutiny has been put to establish the robustness of observational findings and avoid over-interpretation. One long-standing debate is related to cosmological surface brightness dimming, which could introduce a bias against large and diffuse galaxies at high redshift, artificially mimicking an evolution of the mass-size relation (\citealt{Simard+99}; \citealt{Ravindranath+04}). Considerable improvements have been achieved in the last decade, in terms of both data quality and software analysis tools, to keep this effect under control (e.g.\ \citealt{Barden+12}; \citealt{Paulino-Afonso+17}), although the issue is still subject of debate (e.g.\ \citealt{Ribeiro+16}; \citealt{Whitney+20}). A second source of caution is that the conversion between measurable quantities (luminosity and half-light radius) into physical quantities (stellar mass and half-mass radius) is also subject to systematic uncertainties, because (i) semi-empirical color (and color gradient) corrections, needed to bring all the data to a common rest-frame optical wavelength, partly rely on extrapolations in both wavelength and redshift (e.g.\ \citealt{vdW+14}; \citealt{Hill+17}) and (ii) even at a fixed rest-frame optical band, the mass-to-light ratio systematically depends on redshift (as a consequence of galaxy age, e.g.\ \citealt{Barden+05}; \citealt{Trujillo+06}) and galactocentric radius (as a consequence of age gradients, e.g.\ \citealt{Suess+19a}; \citealt{Mosleh+20}). The latter authors in particular have suggested that taking these effects into account may result in significantly slower evolution of the mass-size relation than previously thought, at least for $z \gtrsim 1$, a claim that is awaiting confirmation from the broader community.

The recent time evolution of the mass-size relation of SFGs is also subject of debate, with some studies finding a significant evolution in normalization for $z \lesssim 1$ (e.g.\ \citealt{Mowla+19}\; \citealt{Nedkova+21}) and some others finding little or no evolution out to $z \simeq 1.5$ (e.g.\ \citealt{Nadolny+21}). In addition to the considerations above, uncertainties in the recent evolution may also be related to the fact that, when \emph{directly} comparing observed sizes at different redshifts, a small redshift leverage also implies that variations may become subtle to discern  (e.g. \citealt{Barone+21}). This motivates complementing the classical direct approach with alternative views on the same problem, based on an inherently differential signal of \emph{ongoing} size evolution of SFGs. This approach is explained in more detail in Section \ref{sec:FundamentalConcepts}.

Inevitably, the evolution of the mass-size relation is governed by -- but not synonym to (see Section \ref{sec:FundamentalConcepts}) -- the size evolution of individual galaxies. A number of physical mechanisms can in general contribute to the size evolution of galaxies, also depending on the morphological type. While, for quiescent ellipticals, an important role is believed to be played by mergers (e.g.\ \citealt{Posti+14}; \citealt{vanDokkum+15}), for disc-dominated SFGs there is broad agreement that their evolution is largely dominated by star formation (e.g.\ \citealt{NaabOstriker2006}; \citealt{DuttonvdB2009}; \citealt{Buck+20}) and this view will also be adopted in the rest of this paper (see also further considerations in Section \ref{subsec:discussion_masssize}). Observationally, the way star formation is spatially distributed, within a galaxy, \emph{relative} to the current distribution of stellar mass, provides crucial insight on any \emph{ongoing} size evolution of a SFG (e.g.\ \citealt{MM07}; \citealt{P+15}; \citealt{Nelson+16}; 
\citealt{Wilman+20}). This well established consideration also creates a decisive (perhaps more overlooked) connection with the second topic of this work: the existence and nature of a local main sequence of star formation.

\subsection{Nature of the main sequence of star formation}\label{subsec:intro_mainsequence}
The main sequence of star formation is a relation connecting the stellar content and star formation activity within SFGs. It is known to exist in two flavours: a global (or `integrated') main sequence, relating the total stellar mass $M_\star$ to the total star formation rate (SFR) of a SFG (e.g.\ \citealt{Brinchmann+04}; \citealt{Elbaz+07}; \citealt{Speagle+14}; \citealt{RenziniPeng2015}) and a local (or `spatially resolved') main sequence, relating the surface densities of stellar mass and SFR ($\Sigma_\star$ and $\Sigma_\textrm{SFR}$, respectively) within regions of (typically) $\sim$1 kpc spatial extent (e.g.\ \citealt{Sanchez2020} and references therein). There is broad agreement that these two laws must be physically connected, one relation being an indirect manifestation or a consequence of the other. There are however diverging views about which of the two is the most fundamental and which the corollary.

According to one view, the way galaxies grow is \emph{primarily} governed by large-scale cosmological processes (in particular, the rate of gas accretion onto dark matter haloes), which are then further modulated by internal mechanisms, such as star formation and feedback (e.g.\ \citealt{Bouche+10}; \citealt{Dave+11}; \citealt{Lilly+13}). In this view, the main sequence is fundamentally a global phenomenon. A local relation could however arise, as a consequence of the global one, as long as the radial profiles of gas accretion and star formation are similar for different galaxies (e.g.\ \citealt{Wang+19}). As, however, galaxies are far from perfectly homologous (and accretion profiles are unlikely to be identical), a local main sequence should in this case be, at best, a first-order description, with significant variations expected on a galaxy-by-galaxy basis. 

According to a second view, instead, the local main sequence should be considered the most fundamental (e.g.\ \citealt{Hsieh+17}; \citealt{Enia+20}), possibly indicative of self-regulation of star formation at small-scales (e.g.\ \citealt{Sanchez2020}), while the global main sequence should be regarded as a mere mathematical consequence of the local one upon spatial integration (e.g. \citealt{Cano-Diaz+16}; \citealt{Sanchez+21Mex}). 

Observationally, it should be possible to distinguish between the two scenarios investigating which relation shows the smallest intrinsic scatter (e.g.\ \citealt{Vulcani+19}), or the least dependence on secondary parameters (e.g.\ \citealt{Cano-Diaz+19}). Despite extensive efforts, however, conclusive answers have not been found yet, with different studies coming to opposite conclusions (see e.g.\ recent discussions by \citealt{Ellison+21} and \citealt{Sanchez+21}). Presumably, this is related to differences in sample selection and analysis techniques, in addition to different ways to account for observational uncertainties.

In the meantime that further progress is made from an observational point of view, independent contributions to the debate can also come from investigating the theoretical consequences of one hypothesis or the other on some other aspects of galaxy evolution for which independent constraints are available. This is the approach adopted in the present work. In particular, the consequences will be explored  that the hypothesis of a fundamentally local main sequence would have on the size evolution of individual SFGs -- and thence on the evolution of the mass-size relation.

\subsection{Outline}\label{subsec:outline}

The rest of this work is organised as follows. Section \ref{sec:FundamentalConcepts} summarises the fundamental concepts needed to investigate the evolution of the mass-size relation from a differential perspective. Section \ref{sec:RadialGrowthRate} contains a general derivation of the radial growth rate of a SFG (with potential applications beyond the scope of this paper). In Section \ref{sec:MainSequence}, an application is presented to the particular case in which the main sequence is assumed to be fundamentally local. Implications for the driving questions delineated in this Introduction are discussed in Section \ref{sec:Discussion}. A summary and the main conclusions of this work can be found in Section \ref{sec:Summary}.

A flat $\Lambda$CDM Universe is assumed, with parameters $\Omega_{m,0} = 0.30$ and $H_0 = 70 \; \textrm{km} \; \textrm{s}^{-1} \; \textrm{Mpc}^{-1}$. Conclusions are not affected by changing cosmological parameters within current uncertainties.


\section[Fundamental concepts]{Fundamental concepts}\label{sec:FundamentalConcepts}

\subsection{Specific mass and radial growth rates}\label{subsec:growthrates}
To describe how, at a given point in time, a SFG is `moving' in the mass-size plane, it is useful to consider the (specific) mass growth rate $\nu_\textrm{M}$ and the (specific) radial growth rate $\nu_\textrm{R}$, defined as
\begin{equation}\label{nuMdef}
\nu_\textrm{M} \equiv \frac{1}{M_\star} \frac{d M_\star}{dt} = \frac{d \ln M_\star}{dt} \; ,
\end{equation}
\begin{equation}\label{nuRdef}
\nu_\textrm{R} \equiv \frac{1}{R_{1/2}} \frac{d R_{1/2}}{dt} = \frac{d \ln R_{1/2}}{dt} \; ,
\end{equation}
where $M_\star$ and $R_{1/2}$ are the total stellar mass and the half-mass radius, respectively, and natural logarithms are used.

Note that these positions are similar, but not identical, to those originally introduced by \cite{P+15}, who studied the mass and radial growth rates of the \emph{disc} component of a SFG, in the case that this is well described by an exponential stellar mass density distribution. For a pure exponential disc, the two definitions are equivalent (as, in that case, the half-mass radius is proportional to the disc scale-length). In more general situations, the two definitions are distinct and, in case of ambiguity, the quantities defined in \cite{P+15} could be referred to as $\nu_\textrm{M}^\textrm{disc}$, $\nu_\textrm{R}^\textrm{disc}$. Note that there is no better or worse description in an absolute sense; in particular, the specific radial growth rate can be better defined in terms of the half-mass radius or the disc scale-length, depending on the context and the specific questions being addressed. In this present work, an approach based on $R_{1/2}$ is motivated by the two considerations that (i) the mass-size relation at different redshifts (Section \ref{subsec:intro_masssize}) is most commonly studied in the literature in terms of effective radii (more directly related to the half-mass radius than to the disc scale-length) and (ii) the local main sequence hypothesis (Section \ref{subsec:intro_mainsequence}) does not make \emph{explicit} distinctions between the bulge and disc components of a SFG.

As in \cite{P+15}, the attribute `specific' will be often omitted from now on, for the sake of brevity.\footnote{The attribute `instantaneous' used by \cite{P+15} will also be omitted as partially pleonastic (ideally, a rate is always evaluated at a given time) and strictly speaking not entirely correct (measurable rates are always averaged over some short but finite timescale, dependent on the adopted star formation rate tracer, e.g.\ \citealt{WangLilly20}).} It is also useful to recall that, for a SFG, $\nu_\textrm{M}$ is essentially the same as the specific star formation rate (sSFR), except for corrections due to stellar mass loss, which can for most applications be treated as a normalization rescaling (e.g.\ \citealt{Lilly+13}). While $\nu_M$ is always positive for a SFG, $\nu_R$ can in principle have either sign. A positive radial growth rate ($\nu_\textrm{R} > 0$) is the signature of ongoing inside-out growth.

It is important to emphasize that inside-out growth ($\nu_\textrm{R} > 0$) is \emph{not} a synonym of an evolving mass-size relation. Galaxies can grow inside-out while evolving along a stationary mass-size relation, or along a relation that evolves towards either a higher or lower normalization. The crucial quantity to discriminate between these various situations is not the radial growth rate, but the growth function, as explained below.

\subsection{The growth function and the evolution of the mass-size relation}\label{subsec:growthfunction}

Let a dimensionless parameter, called the \emph{growth function}, be defined as the ratio of the (specific) radial and mass growth rates:
\begin{equation}\label{gdef}
g \equiv \frac{\nu_\textrm{R}}{\nu_\textrm{M}} = \frac{d \log R_{1/2}}{d \log M_\star}\; ,
\end{equation}
where the last equality follows from equations \eqref{nuMdef} and \eqref{nuRdef}, the use of natural or any base logarithm is irrelevant and the derivative is performed, at a given point in time, along the evolutionary path of a SFG in the $(M_\star, R_{1/2})$ parameter space. In this sense, the growth function indicates the \emph{direction} in the mass-size plane towards which a galaxy is moving at the time that the growth function is evaluated. It can also be viewed as the \emph{differential} version of the finite-difference ratio $\Delta \log R_{1/2}/\Delta \log M_\star$, sometimes used to describe evolutionary tracks of galaxies in the mass-size plane (e.g.\ \citealt{vanDokkum+15}; \citealt{Wilman+20}).

The relevance of the growth function for the evolution of the mass-size relation can be appreciated from the following derivation. The mass-size relation can be written in the form
\begin{equation}\label{mass_size}
R_{1/2} = A M_\star^\alpha \; ,
\end{equation}
where $\alpha \simeq 0.22-0.25$, with little or no dependence on redshift (\citealt{Lange+15}; \citealt{vdW+14}; \citealt{Nedkova+21}), while the normalisation $A$ can have some (more or less strong) redshift dependence (see Section \ref{subsec:intro_masssize}). Time derivation of equation \eqref{mass_size} then gives
\begin{equation}\label{derivation1}
\nu_\textrm{R} = \nu_A  + \alpha \nu_\textrm{M} \; ,
\end{equation}
or equivalently
\begin{equation}\label{derivation2}
g = \alpha + \frac{\nu_A}{\nu_\textrm{M}} \; ,
\end{equation}
where equations \eqref{nuMdef}, \eqref{nuRdef} and \eqref{gdef} were used, while
\begin{equation}\label{nuAdef}
\nu_A \equiv \frac{d \ln A}{dt}
\end{equation}
is the (specific) \emph{normalization} growth rate of the mass-size relation.\footnote{A similar derivation was also given by \cite{P+15} in the context of the evolution of exponential discs. That study found $|\nu_A| < 0.01 \; \textrm{Gyr}^{-1}$ for exponential discs in a sample of 35 nearby spiral galaxies.}

The key property of equation \eqref{derivation2} is that there is a \emph{critical growth function}
\begin{equation}\label{equation}
g_\textrm{crit} = \alpha
\end{equation}
for which $\nu_\textrm{A} = 0$. In this case, galaxies grow in size, while they are growing in mass, at the rate that they need in order to move \emph{along} a stationary (non-evolving) mass-size relation.\footnote{In its finite-difference version, this fact has been pointed out multiple times, see e.g.\ \cite{Wilman+20} and references therein.} On the other hand, a growth function larger (smaller) than critical $g > g_\textrm{crit}$ ($g < g_\textrm{crit}$) corresponds to a positive (negative) normalization growth rate $\nu_A > 0$ ($\nu_A < 0$), indicating that galaxies grow in size too fast (too slowly) to remain on the current mass-size relation and the relation itself is evolving towards a higher (lower) normalization at a rate given by $\nu_\textrm{A}$.

Note in particular that a positive evolution of the normalization of the mass-size relation ($\nu_\textrm{A} > 0$) requires that the radial growth rate and the growth function be \emph{not only} positive, but also larger than a critical value ($g > g_\textrm{crit}$). If the growth function is found to be larger than critical, it is then meaningful to compare its value with quantitative predictions of specific evolutionary models. Some examples are given in Section \ref{subsec:predictions}.

\subsection{Evolutionary models}\label{subsec:predictions}
As relevant quantitative examples for the evolution of the mass-size relation, it is useful to consider the two parametrizations explored in the seminal work of \cite{vdW+14}:
\begin{equation}\label{zmodel}
A(z) \; \propto \; (1+z)^{-\beta_z} \quad(\;z\;\textrm{-model}\;)
\end{equation}
and
\begin{equation}\label{Hmodel}
 A(z) \; \propto \; H(z)^{-\beta_H} \quad (\;H\;\textrm{-model}\;) \;.
\end{equation} 
where $H(z)$ indicates the Hubble expansion parameter at redshift $z$. Note that, for convenience, a negative sign is included in the definition of the exponents, so that here $\beta_z$ and $\beta_H$ are defined positive. Using equation \eqref{nuAdef}, these two models translate into the following expressions for the normalization growth rate $\nu_A$:
\begin{equation}
\nu_A = \beta_z H (z) \quad (\;z\;\textrm{-model}\;) \;,
\end{equation}
\begin{equation}
\nu_A = \frac{3}{2} \beta_H \Omega_m(z) H(z) \quad (\;H\;\textrm{-model}\;) \;.
\end{equation}
Note that in both cases the rate of evolution of the mass-size relation decreases towards lower $z$, following the decrease of the Hubble expansion parameter $H$. However, the deceleration effect is more pronounced in the $H$-model, further following the decline in the matter density parameter $\Omega_m$ as a consequence of the take-over of the cosmological constant at recent times. 

For $z=0$ and fiducial cosmological parameters,
\begin{equation}
\nu_A = 5.37 \left( \frac{\beta_z}{0.75} \right) \times 10^{-2} \; \textrm{Gyr}^{-1} \quad (\;z\;\textrm{-model}\;)\;,
\end{equation}
\begin{equation}
\nu_A = 2.15 \left( \frac{\beta_H}{0.67} \right) \times 10^{-2} \; \textrm{Gyr}^{-1} \quad (\;H\;\textrm{-model}\;) \;,
\end{equation}
where the fiducial exponents $\beta_z = 0.75$ and $\beta_H = 2/3$ are the best-fit values found by \cite{vdW+14}.

Replaced in equation \eqref{derivation2} and assuming $\alpha = 0.22$ (e.g. \citealt{Lange+15}) and a typical $\nu_\textrm{M} \sim 0.1 \; \textrm{Gyr}^{-1}$ (e.g.\ \citealt{Speagle+14}; \citealt{McGaugh+17}), these can be translated into expected values for the growth function $g \simeq 0.76$ and $g \simeq 0.43$ for the fiducial $z$-model and $H$-model, respectively. Note that predictions for other choices of the parameters (including e.g.\ at a different redshift) can be readily obtained from the equations above, while slightly different values of $\alpha$ and $\nu_\textrm{M}$ would leave the conclusions of this work unchanged.

\section[Radial growth rate: general case]{The radial growth rate: general derivation}\label{sec:RadialGrowthRate}

To calculate the growth function $g$ (Section \ref{subsec:growthfunction}) and compare it with theoretical expectations (Section \ref{subsec:predictions}), the crucial step is the calculation of the radial growth rate $\nu_\textrm{R}$ (equation \ref{nuRdef}).

A general derivation of $\nu_\textrm{R}$ is readily obtained considering that the half-mass radius  $R_{1/2}$ is defined as the (time dependent) boundary between two (`inner' and `outer') regions of a galaxy with equal stellar mass:
\begin{equation}\label{Rhalfdef}
2 \pi \int_0^{R_{1/2}(t)} R \Sigma_\star(t, R) dR = 2 \pi  \int_{R_{1/2}(t)}^{+ \infty} R \Sigma_\star(t, R) dR \; ,
\end{equation}
where $\Sigma_\star(t, R)$ indicates the azimuthally averaged stellar mass surface density at a galactocentric radius $R$ and time $t$.

It is useful to recall that the time derivative of an integral evaluated over a time dependent domain (of any dimensionality) is
\begin{equation}\label{derivationrule}
\frac{d}{dt} \int_{\Omega(t)} f = \int_{\Omega(t)} \left( \frac{\partial f}{\partial t} + \rm{div} (f\bf{v}) \right) \; ,
\end{equation}
where $\bf{v}$ is \emph{any} regular field that, calculated on the boundary of $\Omega$, gives the velocity of the boundary itself. The time derivative of equation \eqref{Rhalfdef} is then obtained as a simple application of equation \eqref{derivationrule}, for which a convenient valid choice is $\mathrm{v}(R) = \nu_\textrm{R} R$. Simple passages then give
\begin{equation}\label{nuRcalc}
\nu_\textrm{R} = \frac{ \Delta SFR}{4 \pi R_{1/2}^2 \Sigma_{\star, 1/2}} \; ,
\end{equation}
where
\begin{equation}\label{deltaSFR}
\Delta SFR = SFR_\textrm{out}-SFR_\textrm{in} \;,
\end{equation}
is the difference in SFR between the outer an inner half of the galaxy, or explicitly (the time variable being omitted from now on for brevity)
\begin{equation}
SFR_\textrm{in} = 2 \pi \int_0^{R_{1/2}} R \Sigma_{SFR}(R) dR \;,
\end{equation}
\begin{equation}
SFR_\textrm{out} = 2 \pi \int_{R_{1/2}}^{+\infty} R \Sigma_{SFR}(R) dR \;,
\end{equation}
while $\Sigma_{\star,1/2} \equiv \Sigma_\star(R_{1/2})$ and the assumption was made that the function $R^2 \Sigma_\star(R)$ vanishes in both limits $R \to 0$ and $R \to + \infty$, which is always applicable to the distribution of stellar mass in galaxies. It was also assumed that the time derivative of the stellar mass surface density equals the star formation rate surface density $\Sigma_\textrm{SFR}$, as largely appropriate to SFGs (see also Section \ref{subsec:discussion_masssize} below). Following \cite{P+15}, corrections due to stellar mass loss are implicitly incorporated into $\Sigma_\textrm{SFR}$ (which is therefore, strictly speaking, a \emph{reduced} star formation rate surface density). Note that the precise value of the mass loss correction (the `return fraction') is unimportant for the purpose of this work, as it cancels out in the definition of the growth function; the interested reader is referred to \cite{P+15} for further considerations on this topic.

Equation \eqref{nuRcalc} can also be equivalently written as
\begin{equation}\label{nuRalt}
\nu_\textrm{R} = K \Delta (sSFR) \; ,
\end{equation}
where
\begin{equation}
\Delta (sSFR) =  sSFR_\textrm{out}-sSFR_\textrm{in}
\end{equation}
indicates the difference between the (reduced) \emph{specific} star formation rate (sSFR) in the outer and inner half of a galaxy, while
\begin{equation}\label{Kdef}
K \equiv \frac{M_\star}{8 \pi R_{1/2}^2 \Sigma_{\star,1/2}}
\end{equation}
is a dimensionless factor that only depends on the current stellar mass distribution. For instance, $K = 0.475$ for an exponential stellar mass surface density profile and $K = 0.902$ for a de Vaucouleurs profile. A practical expression for the calculation of $K$ for a generic Sersic profile can be found in equation \eqref{Kapprox} below.

Note from equation \eqref{nuRalt} that galaxies with a radially constant sSFR cannot grow in size as a consequence of star formation. A positive radial gradient in sSFR is the necessary and sufficient condition for ongoing inside-out growth, a notion that is well known in the literature (e.g.\ \citealt{MM07}; \citealt{Wang+19} and references therein). The further advantage of equation \eqref{nuRalt} is that it allows to \emph{quantify} the specific radial growth rate associated to a given radial profile of sSFR in a simple but general way. Interestingly, it formally requires only two measurements of the sSFR (the integrated values outside and inside the half-mass radius), so that (with due caution) it could be useful to estimate the radial growth rate of galaxies observed at moderate spatial resolution.

Finally, the growth function $g$ (equation \ref{gdef}) is readily obtained dividing equation \eqref{nuRalt} by the specific mass growth rate $\nu_\textrm{M}$ and recalling that this is the same as the \emph{global} (reduced) sSFR (Section \ref{subsec:growthrates}):
\begin{equation}\label{g_sSFR}
g = K \frac{\Delta (sSFR)}{sSFR} \;.
\end{equation}
The growth function is therefore proportional (through the structural constant $K$) to the \emph{relative} change in sSFR from the inner to the outer half of a galaxy (i.e.\ normalized to the global sSFR).

The formalism and results described above are rather simple and general and could potentially be used for multiple purposes, not necessarily related to the main goals of this paper. In the the rest of this work, however, they will be applied to the specific problem of investigating the implications of a spatially resolved main sequence of star formation for the radial growth rate (and growth function) of SFGs at $z=0$.


\section[Application to a local main sequence]{Application to a spatially resolved main sequence}\label{sec:MainSequence}

\subsection{Resolved main sequence}

The spatially resolved (or local) main sequence of star formation is commonly written in the form
\begin{equation}\label{mainsequence}
\Sigma_\textrm{SFR} = B \Sigma_\star^\beta \; .
\end{equation}
Equivalently, a \emph{normalized} local main sequence can be written in terms of the specific star formation rate:
\begin{equation}\label{normalisedmainsequence}
sSFR = B \Sigma_\star^{-\gamma} \; ,
\end{equation}
with
\begin{equation}\label{gammadef}
\gamma \equiv 1- \beta \; .
\end{equation}

In order to test the hypothesis of a \emph{fundamentally local} main sequence, the assumption will be made (within the scope of this section) that there is some sufficiently small spatial scale (typically, $\sim 1 \; \textrm{kpc}$), where a law such as equation \eqref{mainsequence} or \eqref{normalisedmainsequence} holds, either exactly or in some weaker (e.g.\ azimuthally averaged) sense, but without an \emph{explicit} dependence on galactocentric radius. Further considerations on these conditions and their possible violations will be presented in Section \ref{subsec:discussion_mainsequence}.

Equation \eqref{nuRalt} shows that the condition for a positive radial growth rate (i.e.\ for ongoing inside-out growth) is a radially increasing sSFR. As all galaxies are characterised by a radially declining stellar mass surface densities $\Sigma_\star$, a necessary and sufficient condition for inside-out growth ($\nu_\textrm{R} > 0$), subject to equation \eqref{normalisedmainsequence} and with the assumptions stated above, is that $\gamma>0$ (equivalently, $\beta < 1$). Intuition then suggests that a larger $\gamma$ would qualitatively lead to a larger radial growth rate. This is confirmed and \emph{quantified} in the following.

Evidently, the normalized version of the local main sequence, involving the sSFR (equation \ref{normalisedmainsequence}) is the most  relevant to this study and, accordingly, the slope $\gamma$ that appears in equation \eqref{normalisedmainsequence} will be used as the fundamental parameter. There is however a preference in the observational community to formulate their results in terms of the non-normalized version (equation \ref{mainsequence}). For consistency with these studies, the expression `slope of the main sequence' will be reserved, in the following, for the exponent $\beta$ appearing in equation \eqref{mainsequence}, while the parameter $\gamma$ can then be called either`deviation from unity of the slope of the main sequence' (following equation \ref{gammadef}), or equivalently `slope of the \emph{normalized} main sequence'. Accordingly, the main sequence will be said to be \emph{linear} for $\beta =1$ ($\gamma = 0$), \emph{sub-linear} for $\beta < 1$ ($\gamma > 0$), or \emph{super-linear} for $\beta >1$ ($\gamma < 0$).

Note that both the mass and radial growth rate of a galaxy $\nu_\textrm{M}$ and $\nu_\textrm{R}$ will have a linear dependence on the normalization of the main sequence $B$. However, this dependence cancels out exactly in the calculation of the growth function $g$, which is the ratio of the two (equation \ref{gdef}). Note also that neither the growth rate nor the growth function evaluated at $z=0$ are affected by the fact that the main sequence evolves with redshift (e.g.\ \citealt{Sanchez2020} and references therein). The redshift evolution of the main sequence would be relevant to calculate the \emph{second} time derivative of the half-mass radius with respect to time (or radial growth \emph{acceleration}), which is also straightforward, but not needed for the purpose of this work.

\subsection{Growth function for a resolved main sequence}\label{subsec:growthfunctionmainsequence}

The growth function arising from the assumption of a fundamentally local main sequence is derived here as a special case of the more general formalism described in Section \ref{sec:RadialGrowthRate}.

As shown in Section \ref{sec:RadialGrowthRate}, the calculation of the growth function requires the knowledge or modelling of the current stellar structure of a galaxy. As the focus of this paper is not on individual objects, but on the overall population of SFGs, some parametric form must be adopted. One common choice in similar contexts is to assume an exponential profile (e.g.\ \citealt{MM07}; \citealt{Wilman+20}) and for this reason the exponential case will be discussed in some detail in Section \ref{subsec:exponential}. For the main calculation, however, a more general (though still simple) parametrization will be adopted in the form of a Sersic profile. The choice is dictated by consistency requirements, as this is the most commonly adopted functional form to model the stellar light distribution (and thence derive effective radii) of galaxies in the context of direct determinations of the evolution of the mass-size relation (see Section \ref{subsec:intro_masssize}). It should be stressed, however, that the equations of Section \ref{sec:RadialGrowthRate} can be readily applied to arbitrarily complicated stellar structures, so that the treatment of more elaborate functional forms, or interesting individual cases, is straightforward. An application to the case of two-component models is provided in the Appendix and also serves as a sanity check for the results presented below.

The Sersic profile is given by
\begin{equation}\label{Sersic}
\Sigma_\star (R) = \frac{M_\star}{2 \pi c_n R_{1/2}^2} \exp \left( - b_n \left( \frac{R}{R_{1/2}} \right)^{1/n} \right) \; ,
\end{equation}
where $b_n$ is defined by the condition $P(2n, b_n) = 1/2$, with $P$ the normalised lower incomplete gamma function
\begin{equation}
P(a, x) \equiv \frac{1}{\Gamma(a)} \int_0^x t^{a-1} e^{-t} dt \; ,
\end{equation}
while $c_n = n \Gamma(2n)/b_n^{2n}$. Note also that the structural constant $K$ (equation \ref{Kdef}) is in this case
\begin{equation}\label{KSersic}
K(n) = \frac{c_n e^{b_n}}{4} = \frac{n \Gamma(2n) e^{b_n}}{4 b_n^{2n}} \;.
\end{equation}
Asymptotic expansion (along the lines of \citealt{CiottiBertin1999}) yields the useful approximation
\begin{equation}\label{Kapprox}
K(n) = \frac{\sqrt{\pi n}}{4} \left( 1 + \frac{5}{72 n} + O \left( \frac{1}{n^2} \right) \right) \;,
\end{equation}
which is accurate to 1\% or better for $0.5 < n < 10$.

Assuming a local main sequence with slope $\beta$ (equation \ref{mainsequence}) and inserting equation \eqref{Sersic} in equation \eqref{nuRcalc}, simple passages lead to the result
\begin{equation}\label{mainresult}
g = \frac{\nu_\textrm{R}}{\nu_\textrm{M}} = 4 K(n) \left( \frac{1}{2} - P(2n, \beta b_n) \right) \;.
\end{equation}
Note that, by definition of $b_n$, both the radial growth rate $\nu_\textrm{R}$ and the growth function $g$ vanish  for $\beta = 1$, as expected. Note also that, because $P$ is a cumulative (increasing) function of its second argument, the radial growth rate and the growth function are positive if and only if $\beta < 1$ ($\gamma > 0$), as also expected.

For $\beta$ close to unity, it may be useful to consider the Taylor expansion in the parameter $\gamma = 1-\beta$:
\begin{equation}\label{Taylor}
g = n \left( \gamma + \frac{b_n - 2n +1}{2} \gamma^2 + O(\gamma^3) \right) \; .
\end{equation}
Remarkably, the very simple first order approximation
\begin{equation}\label{approx}
g \simeq \gamma n
\end{equation}
is already accurate to better than 10\% for $1 < n  < 4$ and $ 0 < \gamma < 0.4$ (see Appendix \ref{app:precision} for details). The exact result (equation \ref{mainresult}) will nonetheless be adopted in the following. In particular, the growth function, as calculated from equation \eqref{mainresult}, is shown in Figure \ref{fig:mainresult} and described in some more detail in Section \ref{subsec:implications}, after one short digression on the special case $n=1$ (Section \ref{subsec:exponential}).

\subsection{Comparison with models of exponential discs}\label{subsec:exponential}

The case of an exponential stellar structure ($n=1$) has some particular interest, both for its mathematical simplicity and for being often adopted in the literature. For these reasons, its properties are briefly discussed further here, as a digression. Readers not interested in this special case may move directly to Section \ref{subsec:implications}.

It may be noted that a fundamentally local main sequence (equation \ref{mainsequence}) \emph{implies} that the star formation rate surface density (SFRD) profile of an exponential disc must also be exponential, although with, in general, a different scale-length $R_\textrm{SFR} = \beta^{-1} R_\star$. In particular, the scale length of star formation is larger than that of the stellar mass distribution if and only if the main sequence is sub-linear ($\beta < 1$), which is the condition for inside-out growth.

This is interesting, as a doubly exponential disc (i.e.\ exponential both in stellar mass and SFRD) has often been adopted as an \emph{assumption}, usually for simplicity and regardless of any considerations on the main sequence. Furthermore, a scale-length ratio $R_\textrm{SFR} / R_\star$ larger than unity has long been considered a signature of ongoing inside-out growth (e.g.\ \citealt{MM07}; \citealt{Wilman+20}), in agreement with the condition stated above. However, the conversion of this ratio into a quantitative estimate of the size growth per unit mass growth is usually obtained by means of toy models or heuristic prescriptions.

On the other hand, equation \eqref{mainresult} can be used to derive the growth function of an exponential disc with an exponential SFRD exactly and straightforwardly. Specifically, for $n=1$ equation \eqref{mainresult} reduces to
\begin{equation}\label{exponentialcase}
g =  \frac{e^{b_1}}{b_1^2} \left( (1+\beta b_1)e^{-\beta b_1} - \frac{1}{2} \right) \; ,
\end{equation}
which, taking into account that by definition $2(1+b_1) = e^{b_1}$, can also be written
\begin{equation}\label{exponentialexact}
g = \frac{1}{b_1^2} \left( (1+b_1(1-\gamma))e^{b_1 \gamma} - (1+b_1) \right) \;,
\end{equation}
with $b_1 = 1.678$ and $\gamma = 1-\beta$. Equation \eqref{exponentialexact} also admits the Taylor expansion
\begin{equation}
g = \gamma + \frac{b_1-1}{2} \gamma^2 + O(\gamma^3) \;,
\end{equation}
which can be derived as a special case of equation \eqref{Taylor} or also directly from equation \eqref{exponentialexact}. Numerically, \begin{equation}\label{exponentialTaylor}
g = \gamma + 0.339 \; \gamma^2 + O(\gamma^3) \;.
\end{equation}

The expressions above can be compared for instance with the heuristic prescription proposed by \cite{Wilman+20}, who considered the quantity $\Delta \log R_{1/2}/\Delta \log M_\star$ (the finite-difference version of the growth function, see Section \ref{subsec:growthfunction}) and approximated it as
\begin{equation}\label{W20}
\left( \frac{\Delta \log R_{1/2}}{\Delta \log M_\star} \right)_\textrm{W20} = \ln \frac{R_\textrm{SFR}}{R_\star} = - \ln \beta = \ln \frac{1}{1-\gamma} \;.
\end{equation}

Taylor expansion in $\gamma$ (and comparison with equation \ref{exponentialTaylor}) then shows that the prescription of \cite{Wilman+20} agrees with the exact result at first order in $\gamma$, although deviations appear at second order. As corrections beyond first order are moderate (see Appendix \ref{app:precision}), the calculations above essentially validate the analysis of \cite{Wilman+20}, though limited to the case $n=1$. It may be noted however that there is no special reason to adopt a logarithm in equation \eqref{W20}, as for instance the second-order expansion of equation \eqref{exponentialTaylor}, if not the exact result of equation \eqref{exponentialcase}, would be equally simple and more accurate; in particular, over the range $0 < \gamma < 0.4$, equation \eqref{exponentialTaylor} gives a maximum relative error of 2.5\%, versus 15\% of equation \eqref{W20}.\footnote{Note in passing that \cite{Wilman+20} uses the term \emph{growth factor} to denote the ratio $R_\textrm{SFR}/R_\star$. This shall not be confused with the \emph{growth function} defined in the present work, although for $n=1$ one quantity can be calculated from the other (equation \ref{exponentialcase} with $R_\textrm{SFR}/R_\star = \beta^{-1}$). Note also that \cite{Wilman+20} explicitly explored the impact of stellar mass loss, confirming that it affects the mass and radial growth rates individually, but not their ratio.}

Finally, it is worth emphasising that, because an exponential SFRD for an exponential stellar disc is a \emph{prediction} of equation \eqref{mainsequence}, the existence of exponential discs with a non-exponential SFRD (see e.g.\ \citealt{P+15}) may be regarded as an indication \emph{against} the existence of a fundamentally local main sequence. Further considerations on a non-fundamentally local main sequence can be found in Section \ref{subsec:discussion_mainsequence}.

\subsection{Comparison with theoretical expectations}\label{subsec:implications}

\begin{figure}
\centering
\includegraphics[width=9cm]{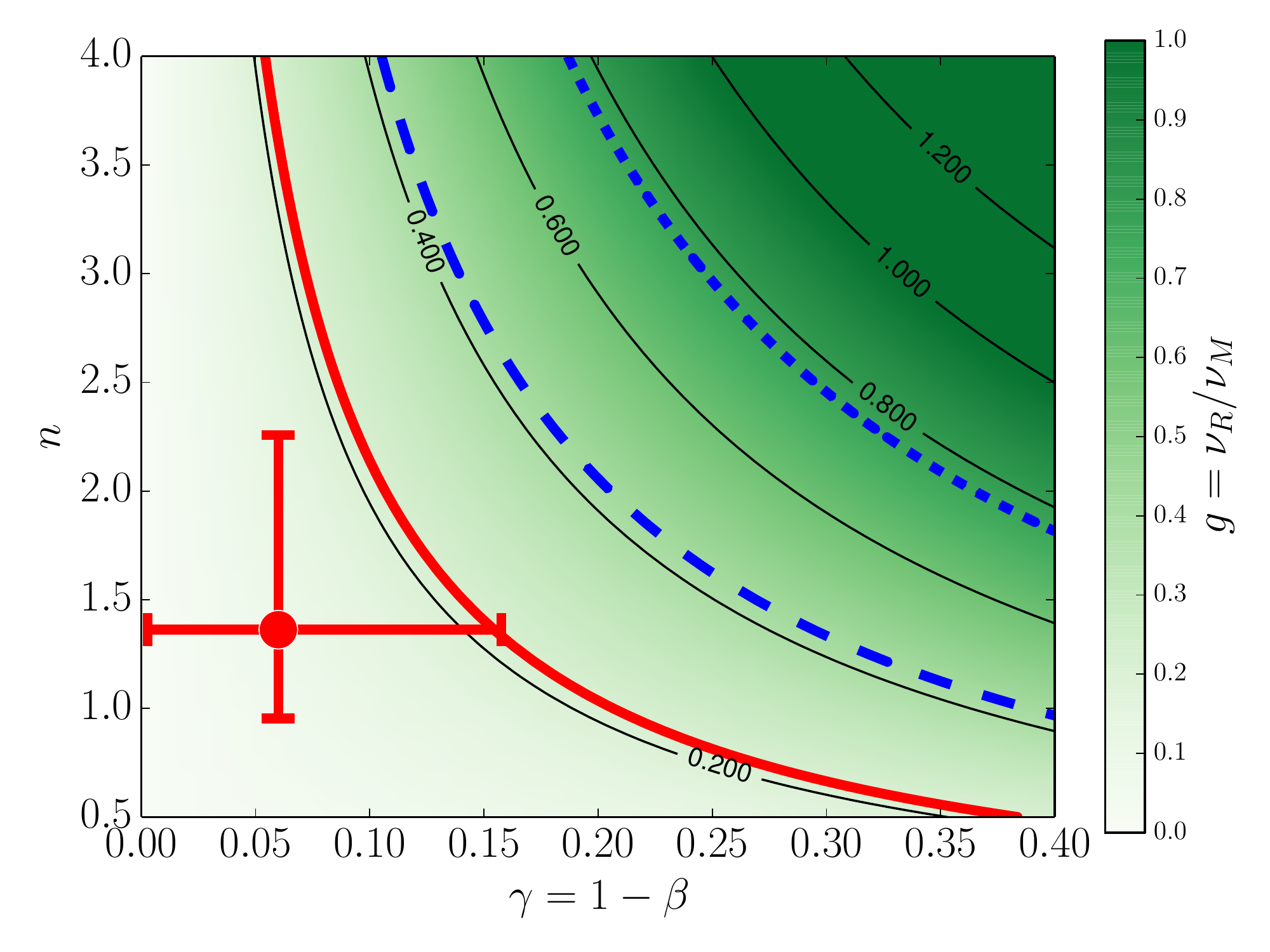}
\caption{Growth function $g \equiv \nu_\textrm{R}/ \nu_\textrm{M} = d \log R_{1/2}/d \log M_\star$ (indicating the \emph{direction} of galaxy evolution in the mass-size plane, see Section \ref{subsec:growthfunction}) as a function of the slope of the \emph{normalized local} main sequence ($sSFR \; \propto \;  \Sigma_\star^{-\gamma}$) and the Sersic index $n$. Contours of constant $g$ are shown as thin solid black lines in intervals of 0.2. Thicker coloured lines indicate, for reference, the \emph{critical} growth function for a stationary mass-size relation (solid red line) and predictions for a mass-size relation that evolves in normalization as $H(z)^{-2/3}$ or $(1+z)^{-0.75}$ (dashed blue line and dotted blue line, respectively). The red point with error bars represents observational constraints for star forming galaxies at $z=0$ (see text for details). A fundamentally local main sequence would drive the evolution of the mass-size relation in a direction opposite to expectations, though marginally consistent with the stationary case.}\label{fig:mainresult}
\end{figure}

To visualise the main result of this work and compare it with expectations, Figure \ref{fig:mainresult} shows the growth function (calculated from equation \ref{mainresult}) as a function of the slope of the normalized main sequence $\gamma = 1-\beta$ (see equations \ref{mainsequence} and \ref{normalisedmainsequence}) and the Sersic index $n$.

As expected, the growth function vanishes for $\gamma=0$ (i.e.\ a perfectly linear main sequence, or $\beta = 1$), independent on the Sersic index. This is because a perfectly linear main sequence implies self-similar evolution and therefore no radial growth. The growth function is instead positive for a sub-linear main sequence ($\beta<1$, $\gamma > 0$) and, for any value of $n$, it becomes larger with increasing deviations from linearity (increasing $\gamma$), as also intuition would suggest. Perhaps less intuitively, for any fixed value of $\gamma$, the growth function increases with increasing Sersic index $n$; this may be understood considering that the susceptibility of a galaxy to radial growth (as measured by the structural constant $K$, see equations \ref{nuRalt} and \ref{Kdef}) increases with $n$ (equations \ref{KSersic}, \ref{Kapprox}).

In addition to contour levels (from 0.2 to 1.2 in steps of 0.2, thin solid black lines), thick coloured lines are overlaid in Figure \ref{fig:mainresult}, as a reference, to highlight the predictions of a few simple scenarios for the \emph{current} evolutionary trend of the mass-size relation, as discussed in Sections \ref{subsec:growthfunction} and \ref{subsec:predictions}. In particular, the solid red line indicates the critical growth function $g_\textrm{crit}$ for a stationary mass-size relation (assuming a slope $\alpha = 0.22$, see Section \ref{subsec:growthfunction}). Note that, as discussed in Section \ref{subsec:growthfunction}, lower (higher) values of $g$ indicate a mass-size relation that is currently evolving towards a lower (higher) normalization; in Figure \ref{fig:mainresult}, this corresponds to the region on the bottom-left (top-right) side of the critical (i.e.\ solid red) line. The dashed and dotted blue lines, located in the super-critical region, indicate the theoretical expectations if the normalization of the mass-size relation is evolving as $H(z)^{-2/3}$ or $(1+z)^{-0.75}$, respectively, i.e.\ following a plain extrapolation to $z= 0$ of the two $z>0$ reference models by \cite{vdW+14} (see Section \ref{subsec:predictions}).

To visualise observational constraints on the slope of the local main sequence, a useful aid is provided by the compilation of \cite{Enia+20} (see their Table 2). This is chosen because it encompasses a large variety of sample definitions, star formation rate indicators and analysis strategies. It is important to emphasize that all these factors are subject of active discussion in the observational community and cannot be fully captured by a statistical description. This prevents, for instance, drawing conclusions in terms of confidence levels. Nonetheless, Figure \ref{fig:mainresult} includes a formal representation of the mentioned constraints, with the red point position and bar in the horizontal direction representing, respectively, the median and the 16th-84th percentile range of $\gamma$, as derived from the compilation of \cite{Enia+20}. More considerations about observational determinations of $\gamma$, including outliers, can be found in Section \ref{subsec:discussion_mainsequence}. 

Similarly, the point position and bar in the vertical direction indicate, respectively, the median and the 16th-84th percentile range of the Sersic index of late-type galaxies ($0 < T < 10$ according to the classification of \citealt{Buta+15}) at $z=0$, as derived from the Spitzer Survey of Stellar Structure in Galaxies (S$^4$G, \citealt{Sheth+10}), following the analysis of \cite{Salo+15}. This is a rather robust constraint, consistent with other independent determinations in the nearby Universe (e.g.\ \citealt{Lange+15}). Note also that the results shown here remain unchanged if a two-component model is adopted instead of a single Sersic profile (see Appendix \ref{app:2comp} for details).

The comparison in Figure \ref{fig:mainresult} clearly shows that, within the constraints described above, the hypothesis of a fundamentally local main sequence would imply relatively small values of the growth function, preferentially lower than (but marginally consistent with) the critical value $g_\textrm{crit}$ for a stationary mass-size relation. On the other hand, a fundamentally local main sequence appears inconsistent with a significant ongoing positive evolution of the normalization of the mass-size relation of SFGs. In particular, it appears in tension with the extrapolation at $z=0$ of both the parametric reference models described in Section \ref{subsec:predictions}.

Further discussion of this finding, including possible ways to resolve the tension and possible implications for each of the conflicting elements, is the subject of Section \ref{sec:Discussion}.

\section[Discussion]{Discussion}\label{sec:Discussion}

It was found in Section \ref{subsec:implications} that the assumption of a fundamentally local main sequence of star formation is in tension with the assumption of a strong ongoing evolution of the mass-size relation of SFGs in the direction of an increasing normalization.

Formally, this result does not indicate which of the two hypotheses is false (and does not exclude that both of them may be), indicating however that they cannot both be true. Two complementary questions therefore arise and are briefly discussed below: first (Section \ref{subsec:discussion_masssize}) whether the mass-size relation of SFGs is currently undergoing significant evolution, second (Section \ref{subsec:discussion_mainsequence}) whether the main sequence of star formation is fundamentally local.

\subsection{Is the mass-size relation evolving with time?}\label{subsec:discussion_masssize}

If a local main sequence is a good description of star formation in SFGs at $z \sim 0$, then the analysis in Section \ref{subsec:implications} indicates a preference for a sub-critical growth function, marginally consistent with critical, $g \lesssim g_\textrm{crit}$ (Figure \ref{fig:mainresult}). A strictly subcritical value $g < g_\textrm{crit}$ would imply a mass-size relation that is currently evolving towards a \emph{lower} normalization (see Section \ref{subsec:growthfunction}). This is at odds with theoretical expectations and other observational indications (see Section \ref{subsec:intro_masssize}) and should at least in first instance be considered unlikely. 

A close-to-critical value of the growth function $g \simeq g_\textrm{crit}$, on the other hand, is in agreement with a number of other studies, based on a finite-difference version of the growth function, or\ $\Delta \log R_{1/2} / \Delta \log M_\star$ (see Section \ref{subsec:growthfunction}). \cite{vanDokkum+13} estimated  $\Delta \log R_{1/2} / \Delta \log M_\star \sim  0.27 \simeq g_\textrm{crit}$, between $z \simeq 2.5 $ and $z \simeq 0$, for the evolution of Milky-Way-like galaxies under the assumption of evolution at fixed cumulative number density. This is in good agreement with $\Delta \log R_{1/2} / \Delta \log M_\star \sim  0.26$ found by \cite{Wilman+20} comparing H$\alpha$-based SFRD scale-lengths to stellar mass scale-lengths of $0.7 < z < 2.5$ SFGs (see also Section \ref{subsec:exponential}). These findings are consistent with SFGs moving along a close to stationary mass-size relation, possibly since as early as $z \simeq 2$. They also agree with inference from spectro-photometric and chemical evolution modelling of $z=0$ galaxies (\citealt{MM11}) and a differential estimate of the growth function of nearby discs (\citealt{P+15}). At least at first sight, however, all these  results are in disagreement with observational indications for a strongly evolving normalization of the mass-size relation of SFGs, based on the direct comparison of rest-frame optical effective radii, as a function of stellar mass and redshift (see references in Section \ref{subsec:intro_masssize}).

The contradiction may have a physical explanation. For instance, \cite{Wilman+20}, who also clearly pointed out the problem, speculated that it could be resolved invoking `aggressive' selective quenching of compact SFGs, or an important contribution of minor mergers to the evolution of SFGs between $z = 1$ and $z = 0$. The first idea was drawn qualitatively along the lines of the investigation of \cite{AbramsonMorishita2018}, but quantitative consistency was not investigated and remains unclear.\footnote{It may be noted here that the `null hypothesis' of \cite{AbramsonMorishita2018} (individual galaxies evolve at constant effective surface density) would imply a growth function $g = 0.5$, almost double than $\Delta \log(R_{1/2})/\Delta \log (M_\star) = 0.26$ found by \cite{Wilman+20}; therefore, in the language of \cite{AbramsonMorishita2018}, the results of \cite{Wilman+20} may in fact be better regarded as a `falsification of the null hypothesis'.} The subject of minor mergers requires some distinctions. Wet (gas-rich) mergers could contribute bringing fresh gas to fuel star formation, though they are probably \emph{not} the main contributors to gas accretion (\citealt{DiTeodoro+14}); this effect would in all cases be counted as star-formation-induced growth and therefore not help to solve the problem. On the other hand, dry (gas-poor) mergers constitute a source of dissipationless stellar accretion from (almost) random orientations and can therefore give relevant contributions to the growth of spheroidal (dispersion-supported) structures, such as stellar haloes (e.g.\ \citealt{Helmi2008}; \citealt{Das+16}; \citealt{Helmi2020}) or elliptical galaxies (e.g.\ \citealt{Newman+12}; \citealt{Posti+14}; \citealt{vanDokkum+15}), but hardly to the main body of rotationally-supported SFGs. Another potentially relevant physical process is stellar radial migration. Although important in terms of mixing (\citealt{SB02}, \citealt{SB09}), however, the dominant mode of radial migration is associated to little dynamical heating and, as a consequence, to minor or slow changes in the overall stellar structure of a galaxy (\citealt{Sellwood2014}; \citealt{Frankel+20}). In particular, theoretical studies including both star formation and stellar radial migration confirm that star-formation is the dominant channel for the structural growth of SFGs (e.g.\ \citealt{MCM14}; \citealt{Frankel+20}; \citealt{Buck+20}). None of the elements listed so far therefore appears to provide a compelling solution.

Possibile explanations of the paradox could also be (and have been) looked for in systematic biases in the observed evolution of the mass-size relation, such as those related to cosmological surface brightness dimming, or the conversion of half-light radii into half-mass radii (see Section \ref{subsec:intro_masssize}). Among recent quantitative indications in this direction, \cite{Suess+19a} and \cite{Mosleh+20} have argued that modelling the effect of radial gradients in (rest-frame optical) mass-to-light ratios results in a slower evolution of the mass-size relation compared to what inferred based on half-light radii, at least for $1 < z < 2$. This scenario would imply $g \simeq g_\textrm{crit}$ for $1 <z < 2$. However, it is not clear whether it could account for  $g \simeq g_\textrm{crit}$ between $z =1$ and $z = 0$, as the evidence discussed earlier would indicate. In particular, \cite{Suess+19b} found that SFGs with $M_\star = 10^{10.5} \; \textrm{M}_\odot$ have an approximately constant half-mass radius $R_{1/2} \sim 3 \; \textrm{kpc}$ between $z=2$ and $z=1$. However, to then reach the observed sizes of SFGs with this mass at $z=0$ ($R_{1/2} \simeq 4.5 \; \textrm{kpc}$, e.g.\ \citealt{Lange+15}), a further factor 1.5 increase in normalization is still required between $z = 1$ and $z = 0$. This is similar to the prediction of a $z$-model with $\beta_z = 0.75$ (equation \ref{zmodel} and dotted line in Figure \ref{fig:mainresult}) and inconsistent with $g \simeq g_\textrm{crit}$ at $z=0$. Though appealing, therefore, also this solution does not appear completely satisfactory at the present time, unless some additional systematic effect is identified.\footnote{It is worth mentioning that, as convincingly shown by \cite{Frankel+19}, despite not providing an appreciable \emph{physical} contribution to radial growth, the \emph{mixing} effect of radial migration can blur the \emph{observational} signature of inside-out growth and hinder its accurate reconstruction from spatially resolved stellar populations of a finite age, as well as from their proxies, such as colour gradients.}
 
Finally, one (somewhat complementary) possibility could be that the mass-size relation may have evolved strongly during some early epochs, but then reached an equilibrium state at some finite (to be determined) look-back time. Retrospectively, there have been a number of indications in favour of this scenario in the literature. One way of illustrating this is to consider that the simplest phenomenological parametric description $A(z) \propto (1+z)^{-\beta_z}$ (the $z$-model in Section \ref{subsec:predictions}) has generally been providing systematically different values of $\beta_z$, depending on the adopted redshift baseline for its measurement. Studies focused on the evolution at $z > 1$ tend to find a relatively large $\beta_z \sim 1$ (e.g.\ \citealt{Oesch+10}; \citealt{Allen+17}; \citealt{Yang+21}). On the other hand, investigations focused at $z \lesssim 1$ have consistently been finding considerably smaller values. Extending the pioneering work of \cite{Lilly+98} and including a treatment to convert surface brightness into mass surface density, 
\cite{Barden+05} found evidence for a non-evolving mass-size relation ($\beta_z = 0$) since $z =1$ (though note that \citealt{vdW+14} associates a small but non-vanishing $\beta_z = 0.2$ to the same work). A similarly small value $\beta_z = 0.17$ can be derived from the more recent results of \cite{Barone+21} in the redshift range $0.014< z < 0.76$, while \cite{Nadolny+21} found statistically insignificant evolution since $z \sim 1.5$ (see also \citealt{Nagy+11}, who found $\beta_z = 1.42$ for $z > 1.5$ and predicted convergence to the local relation by $z \sim 1$). Perhaps unsurprisingly, studies involving intermediate baselines $0.2 \lesssim z \lesssim 2.5$ find intermediate values $\beta_z \sim 0.75$ (\citealt{vdW+14}; \citealt{Paulino-Afonso+17}). This could indicate that one single, perhaps intermediate, value of $\beta_z$ is a good description at all redshifts (e.g.\ \citealt{LillyCarollo2016}; \citealt{Renzini2020}), but it could also more prosaically indicate the average value of an intrinsically varying quantity. The second interpretation appears favoured by the analysis of \cite{vdW+14}, who in fact introduced the alternative parametrization in terms of the Hubble parameter (the $H$-model in Section \ref{subsec:predictions}) to partially encapsulate this effect and attenuate the drawback of otherwise overpredicting galaxy sizes at $z = 0$.  As expected, the $H$-model was indeed found to be an improvement over the $z$-model in this respect (also in agreement with Figure \ref{fig:mainresult} of this work). Nonetheless, \cite{vdW+14} also showed that the \emph{recent} evolution may be slower than any of the two descriptions would predict and in fact consistent with the earlier results of \cite{Barden+05}.

In conclusion, it appears possible that a currently stationary mass-size relation is not in real tension with independent observational constraints, with apparent discrepancies originating not (or at least not only) in the measurements themselves, but (at least in part) in their parametrization. It should not be excluded, however, that other factors, such as the issue of converting effective radii into half-mass radii, or other systematic effects, may be playing some role as well. If this interpretation is correct, it could be meaningful to shift part of our attention from the common question: `What exponent (or simple functional form) describes the evolution of the mass-size relation across all cosmic times?' in favour of another question: `When and how did the mass-size relation settle into a stationary (or almost stationary) state?'.

From the vantage point of this present work, a possible way to determine the moment of this transition could be to identify the closest look-back time for which a spatially resolved main sequence starts showing substantial deviations from linearity (sufficiently large $\gamma$, see Section \ref{sec:MainSequence}). Note that a spatially resolved main sequence has been proposed to be in place already at $z \simeq 1$ (\citealt{Wuyts+13}), with a higher normalization compared to $z=0$ (as expected), but similarly modest deviations form linearity ($\gamma = 0.05-0.16$, depending on the fit domain). This is in further agreement with the idea that the mass-size relation was already growing slowly (or not at all) at $z\simeq1$. However, drawing strong conclusions from the spatially resolved main sequence at $z\simeq1$ may still require some caution, considering that already at $z=0$ there are still not entirely settled open questions on the topic. This is further discussed in the next section.

\subsection{Is the main sequence fundamentally local?}\label{subsec:discussion_mainsequence}

Complementary to what discussed in Section \ref{subsec:discussion_masssize}, the formal disagreement between observational constraints on the local main sequence and popular parametric models for the size evolution of SFGs (Figure \ref{fig:mainresult}) could also be considered a challenge for the local main sequence framework.

Before delving into this interpretation, it should once again be emphasized that, although broadly representative of the current literature, the (formally $1 \sigma$) horizontal range indicated in Figure \ref{fig:mainresult} should be considered with caution, as variations between different main sequence determinations are in large part methodological are therefore dominated not by random errors but by systematics. This makes it worth to at least briefly discuss outliers. Among these, it's worth mentioning \cite{Cano-Diaz+16}, who found $\beta = 0.72$ ($\gamma = 0.28$). For a Sersic index $n=1.5$, this would give a growth factor $g=0.42$, formally in good agreement with the model $A(z) \; \propto \; H(z)^{-2/3}$. However, the discrepancy with other works has been convincingly explained in terms of a sub-optimal fitting methodology (\citealt{Hsieh+17}), a conclusion also confirmed in a later work by \cite{Cano-Diaz+19}, who found $\beta = 0.94$, corresponding to $\gamma = 0.06$ and  $g \sim 0.1 < g_\textrm{crit}$, in agreement with the rest of the literature. On the opposite end, there have been occasional reports of a super-linear local main sequence ($\beta > 1$, $\gamma < 0$, e.g.\ \citealt{Vulcani+19}; \citealt{Ellison+21}). This would formally correspond to a radially declining sSFR, in contrast with independent evidence that sSFR gradients in SFGs are generally positive (e.g.\ \citealt{MM07}; \citealt{Gonzalez-Delgado+16}; \citealt{Wang+19}). Super-linear slopes should however be interpreted with caution, as they are found together with a considerable scatter and, reportedly, do not apply to individual galaxies (see also a related discussion below).

If one accepts the compilation of \cite{Enia+20} as representative, then the discrepancy with simple models of the evolution of the mass-size relation is real. One possible solution could involve relaxing the assumption that the local main sequence is a fundamental law, perhaps in favour of a scenario in which the main sequence is instead fundamentally global. As mentioned in Section \ref{subsec:intro_mainsequence}, indications in this direction could come from investigations of the scatter of the local main sequence, as well as the possible existence of secondary dependences. In this context, \cite{Vulcani+19} found that the local main sequence has a larger scatter than the global one, which could suggest that the latter is more fundamental (although \citealt{Vulcani+19} did not come to this conclusion). Furthermore, the scatter of the local main sequence has been proposed to correlate strongly with 
galaxy morphology (\citealt{Gonzalez-Delgado+16}; \citealt{Cano-Diaz+19}), but only weakly with stellar mass (\citealt{Erroz-Ferrer+19}; \citealt{Sanchez2020}). This is somewhat puzzling, given that morphology and stellar mass are known to be correlated, and could be indicative of inconsistencies between different studies. Finally, \cite{Ellison+21} found that the scatter of the local main sequence correlates with that of other known relations, in particular those involving the molecular gas surface density, and concluded that the local main sequence is not a fundamental law. However, \cite{Sanchez+21} has disputed this argument and proposed that the correlations observed by \cite{Ellison+21} can be explained in terms of observational uncertainties, concluding that the local main sequence is in fact fundamental.

While the disagreement between observational studies is waiting to be composed, it is important to realise that not every systematic variation of the local main sequence would be sufficient to reconcile it with a size evolution as fast as expected. In particular, systematic variations \emph{in the normalization} (the focus of most of the investigations mentioned above) would not serve to alleviate the problem, 
as SFGs obeying local main sequences with a varying normalization but a common slope would evolve in the mass-size plane at a different pace but all in the wrong direction. Interestingly in this context, there have also been reports of a systematic dependence of the slope of the local main sequence on morphology. In particular, the results of \cite{Maragkoudakis+17} indicate a variation from $\gamma = 0.03$ for late-type spirals to $\gamma = 0.19$ for early-type spirals. Combined with a sufficiently large Sersic index ($n \geq 2$), this could in principle bring the growth function in agreement with the $H(z)^{-2/3}$ model for the earliest type spirals. The issue with late-type spirals, however, would remain unsolved.

One formal solution could be that individual galaxies obey a local main sequence that is sufficiently shallow (sufficiently large $\gamma$) to allow for fast radial growth, but, combined to the effect of a varying normalization, they appear to lie collectively on an \emph{ensemble} relation that is steeper. The results of \cite{Vulcani+19} seem to support this scenario, as their local main sequence fits for \emph{individual} galaxies is significantly shallower than the ensemble. On the other hand, \cite{Enia+20} have found instead that individual galaxies obey a local main sequence with a slope consistent with that of the ensemble, so once again observational determinations appear in conflict with each other. One key statistical test can be devised for the mentioned scenario: in fact, if a galaxy is followed along its hypothetical relatively shallow track in the $(\Sigma_\star, \Sigma_\textrm{SFR})$ plane, it should cross the ensemble relation (from relatively high to relatively low sSFR) with increasing surface density $\Sigma_\star$ and decreasing galactocentric radius $R$. This should be detectable, at a statistical level, in the form of anti-correlation between residual sSFR (deviation from the ensemble relation) and galactocentric radius. However, \cite{Hall+18} investigated the possible existence of such a correlation and found none.

Finally, one possible reason why the main sequence is found to be closer to linear compared to some expectations may be related to selection effects. Some studies (e.g.\ \citealt{Cano-Diaz+19}) define the local main sequence only in `active' regions, separated from `retired' regions on the basis of some SFRD threshold. Including all regions would inevitably contribute some scatter, but also a decrease of the median $\Sigma_\textrm{SFR}$ at fixed $\Sigma_\star$. If the fraction of `retired' regions was found to increase with increasing surface density, it would eventually induce a shallower slope, perhaps reducing the discrepancy with theoretical expectations.\footnote{This is formally similar to (but physically distinct from) an effect reported by \cite{Pessa+21} in terms of `detection fraction', which however originates in \emph{low} surface density regions and appears related to sensitivity effects and the small probed spatial scales (100 pc).} Although a dedicated quantitative analysis (e.g.\ in terms of `retired fraction') is still lacking, a similar trend is at least qualitatively visible for the earliest-type spirals (\citealt{Medling+18}) and the most massive SFGs ($M_\star > 10^{10.5} \; \textrm{M}_\odot$, \citealt{Wang+19}). For lower masses and later types, however, `retired' regions appear to be rare and with a plausibly minor impact on the slope of the main sequence. This is also consistent with the fact that a number of studies (e.g.\ \citealt{Hall+18}; \citealt{Enia+20}) find a close to linear relation without introducing distinctions between regions, further suggesting that selection effects are probably not sufficient to change the picture dramatically. At the same time, it should be noted that if a local main sequence was to apply only to some regions and not to others, the mathematical link between the global and local main sequence would be formally broken, as the `retired' regions would contribute to the global relation (in terms of total stellar mass), but not to the local one. In particular, this would be an obstacle to the idea that the global main sequence is a consequence of the local one upon spatial integration.\footnote{This fact was explicitly acknowledged as a limitation by \cite{Cano-Diaz+16}, while \cite{Sanchez2020} suggested it could help explaining the shallower slope of the \emph{global} main sequence.}

In summary, there are a number of indications suggesting that the local main sequence could be a not truly fundamental law. Nonetheless, none of the considered proposed deviations from universality appears clearly sufficient to evade the implications presented in Section \ref{sec:MainSequence}. As a consequence, the spatially resolved main sequence, even if not necessarily fundamental, arguably remains a useful tool for the investigation of the size evolution of SFGs. Further studies, especially if directed at reducing or explaining some of the remaining discrepancies in the observational literature, would be valuable to consolidate, or challenge, the conclusions of this work.


\section[Summary and conclusions]{Summary and conclusions}\label{sec:Summary}

The rest-frame optical effective radii of star forming galaxies (SFGs) appear to be increasing, at fixed stellar mass, with decreasing redshift; however, it is still debated whether the same holds true for half-mass radii (which would imply an evolution of the physical mass-size relation of SFGs) and, furthermore, it is not clear whether one single parametric law can be used to describe such evolution across all cosmic times. Apparently unrelated, a main sequence of star formation is observed to hold for SFGs both at local and global scales and it is not clear which of the two should be considered the most fundamental. 

In this work, a strong connection between these two seemingly distant problems is shown to exist, in the sense that making hypotheses on one of the two has clear implications on the other, so that not every pair of assumptions is a viable possibility. The salient points of this investigation can be summarised as follows.
\begin{enumerate}
\item{The concepts of specific mass growth rate ($\nu_\textrm{M} \equiv \dot{M}_\star/M_\star$) and specific radial growth rate ($\nu_\textrm{R} \equiv \dot{R}_{1/2}/R_{1/2}$), analogous to those introduced by previous work in the context of exponential discs, are generalised to an arbitrary stellar mass distribution. The \emph{growth function} $g \equiv \nu_\textrm{R}/\nu_\textrm{M}$, defined as the ratio of the two, can be used to identify the direction of evolution of SFGs in the mass-size plane. A growth function larger than a critical value, or $g > g_\textrm{crit} = \alpha$, where $\alpha$ is the slope of the mass-size relation, is required for a relation that is evolving in the direction of an increasing normalization.}
\item{Exact expressions are derived for the specific radial growth rate $\nu_\textrm{R}$ and the growth function $g$ of a SFG given \emph{any} known distribution of stellar mass and star formation rate surface densities. A number of equivalent expressions are provided in Section \ref{sec:RadialGrowthRate}. Notably, the specific radial growth rate is found to be $\nu_\textrm{R} = K \Delta(sSFR)$, where $\Delta(sSFR)$ is the difference in reduced sSFR between the outer and inner half of a galaxy (as defined by the half-mass radius) and $K$ is a dimensionless structural constant (equation \ref{Kdef}). A practical formula for the calculation of $K$ is provided in equation \eqref{Kapprox}. With some caution, the expression above could be used to measure the specific radial growth rate of distant galaxies with a limited spatial resolution. This result is independent on any assumptions on the local main sequence.}
\item{For the particular case of a fundametally local main sequence, an expression is derived for the growth function, in terms of the slope of the \emph{normalized} main sequence $\gamma$ (such that $sSFR \; \propto \; \Sigma_\star^{-\gamma}$, or equivalently $\Sigma_\textrm{SFR} \; \propto \; \Sigma_\star^{1-\gamma}$) and the Sersic index $n$ of a galaxy. An exact result is provided in terms of the incomplete gamma function (equation \ref{mainresult}). However, the remarkably simple approximation $g \simeq \gamma n$ is found to be accurate to better than 10\% for $1<n<4$ and $0 < \gamma < 0.4$ and could be of practical use for future studies, unless a higher precision is required.}
\item{Under the assumption of a fundamentally local main sequence, the growth function for $z=0$ SFGs is found to be smaller than, or at most marginally consistent with, the critical value for a \emph{stationary} mass-size relation. It is instead inconsistent with the growth function predicted by simple parametrizations of the evolution of the normalization $A$ of the mass-size relation, in terms of either the scale factor $A(z) \; \propto \; (1+z)^{-0.75}$ or the Hubble expansion parameter, $A(z) \; \propto \; H(z)^{-2/3}$.}
\end{enumerate}

The last result in particular implies that the two hypotheses of a \emph{fundamentally local} main sequence and \emph{ongoing} evolution of the mass-size relation (according to the simple parametrizations above) are mutually inconsistent, indicating that \emph{at least one of the two must be incorrect}.

The methods presented in this work cannot determine which of the two hypotheses should be discarded. However, arguments were offered in the discussion that this may in fact be the case for both.  As an educated guess, a scenario is delineated where the spatially resolved main sequence is not a truly fundamental law, but phenomenologically good enough to provide meaningful constraints on the size evolution of SFGs. On the other hand, the mass-size relation of SFGs may have converged to a stationary state at some finite lookback time, in agreement with significant, though partly indirect, independent evidence. Disagreement with other conflicting evidence may be partly due to systematic effects (e.g.\ related to the conversion of light into mass) but also in part be only apparent and indicative that simple parametrizations in terms of $(1+z)$ or $H(z)$ do not correctly capture the evolution at high $z$ and low $z$ simultaneously.

Observational investigations on the properties of a spatially resolved main sequence at $z > 0$ could contribute to determine the moment in the past when the transition to a stationary state occurred. Current indications suggest that this event may have happened at $z \gtrsim 1$.

\section*{Acknowledgements}
I would like to thank Alvio Renzini and Simon Lilly for stimulating discussions that have contributed to motivate this study. This work was supported by the Netherlands Research School for Astronomy (NOVA).

\section*{Data Availability}
No new data were generated or analysed in support of this research.

\bibliographystyle{mnras}
\bibliography{mybib}


\appendix
\section[Appendix A]{Precision of the {\Large{\lowercase{$g = \gamma n}$}} approximation}\label{app:precision}
\begin{figure}
\centering
\includegraphics[width=9cm]{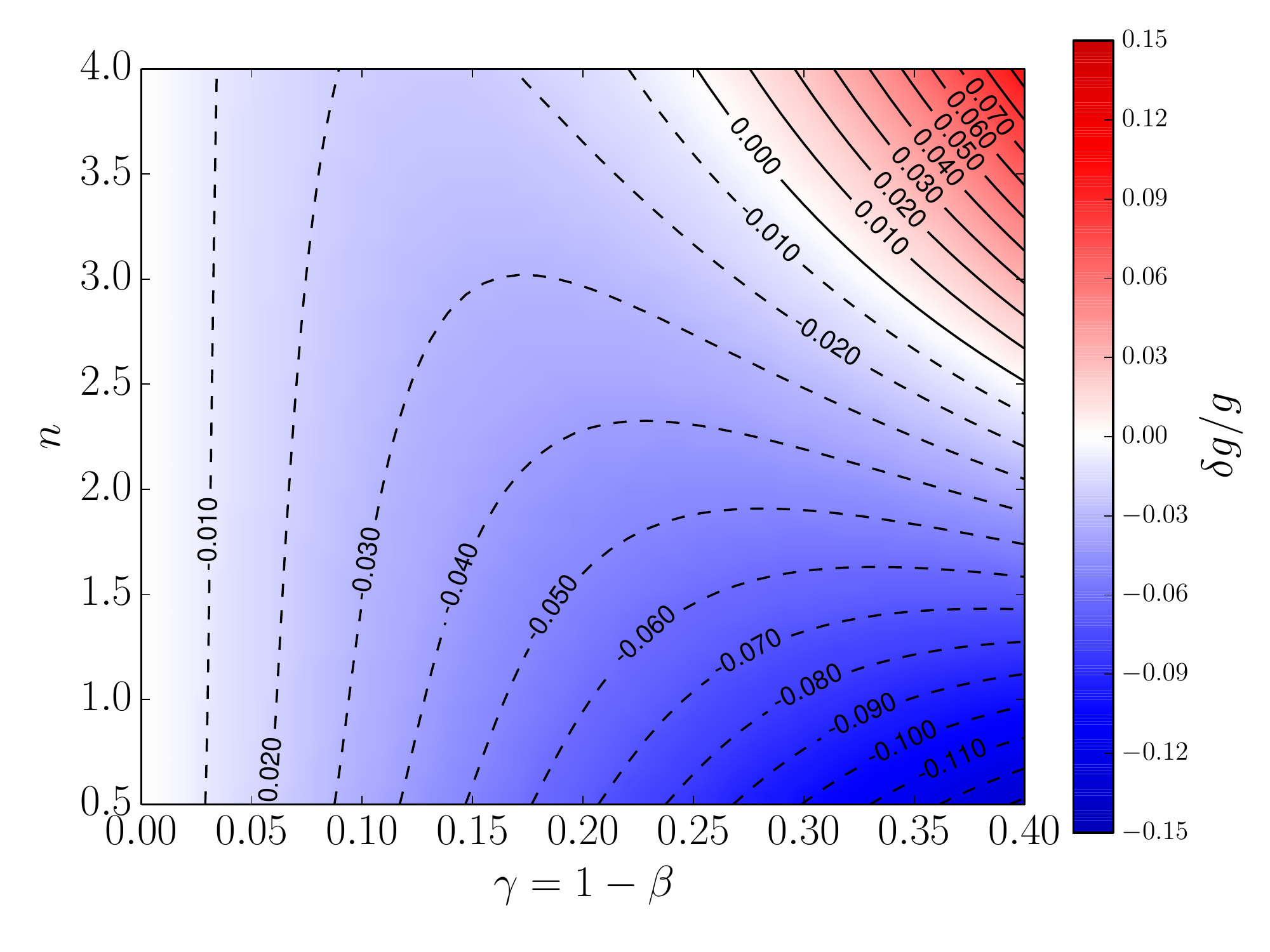}
\caption{Relative error $\delta g /g$ associated to the approximation $g \simeq \gamma n$ to the growth function, compared to the exact expression (equation \ref{mainresult}). Variables on the axes are as in Figure \ref{fig:mainresult}.}\label{fig:approx}
\end{figure}
Figure \ref{fig:approx} provides a visual representation of the relative error associated to the simple approximation to the growth function $g \simeq \gamma n$, compared to the exact result provided in equation \eqref{mainresult}. The relative error is expressed as $\delta g/g= (g_\textrm{approx}-g_\textrm{exact})/g_\textrm{exact}$, where $g_\textrm{exact}$ and $g_\textrm{approx}$ are the exact and approximate expression for $g$, respectively. For most of the parameter space, the simple approximation $g \simeq \gamma n$ underestimates the true value of $g$ by just a few per cent. It instead provides an overestimate, again by only a few per cent, when both $\gamma$ and $n$ are sufficiently large, as indicated. The absolute value of the relative error remains smaller than 10\% across $0 < \gamma < 0.4$ and $1 < n < 4$. A slightly larger (and negative) relative error (up to 13\%) can be found for $\gamma > 0.3$ and $0.5 < n < 1$. Note that the main analysis in this paper was obtained using the exact expression for the growth function. The approximate version would however leave the results unchanged and could also be used by future studies on related topics, unless a better precision than indicated in Figure \ref{fig:approx} is required.

\section[Appendix B]{Two-component models}\label{app:2comp}

As a sanity check, it is investigated here whether the results of Section \ref{subsec:implications} would be altered by adopting a two-component (bulge+disc) description for the stellar mass surface density distribution of SFGs, instead of a single Sersic profile.

\begin{figure}
\centering
\includegraphics[width=9cm]{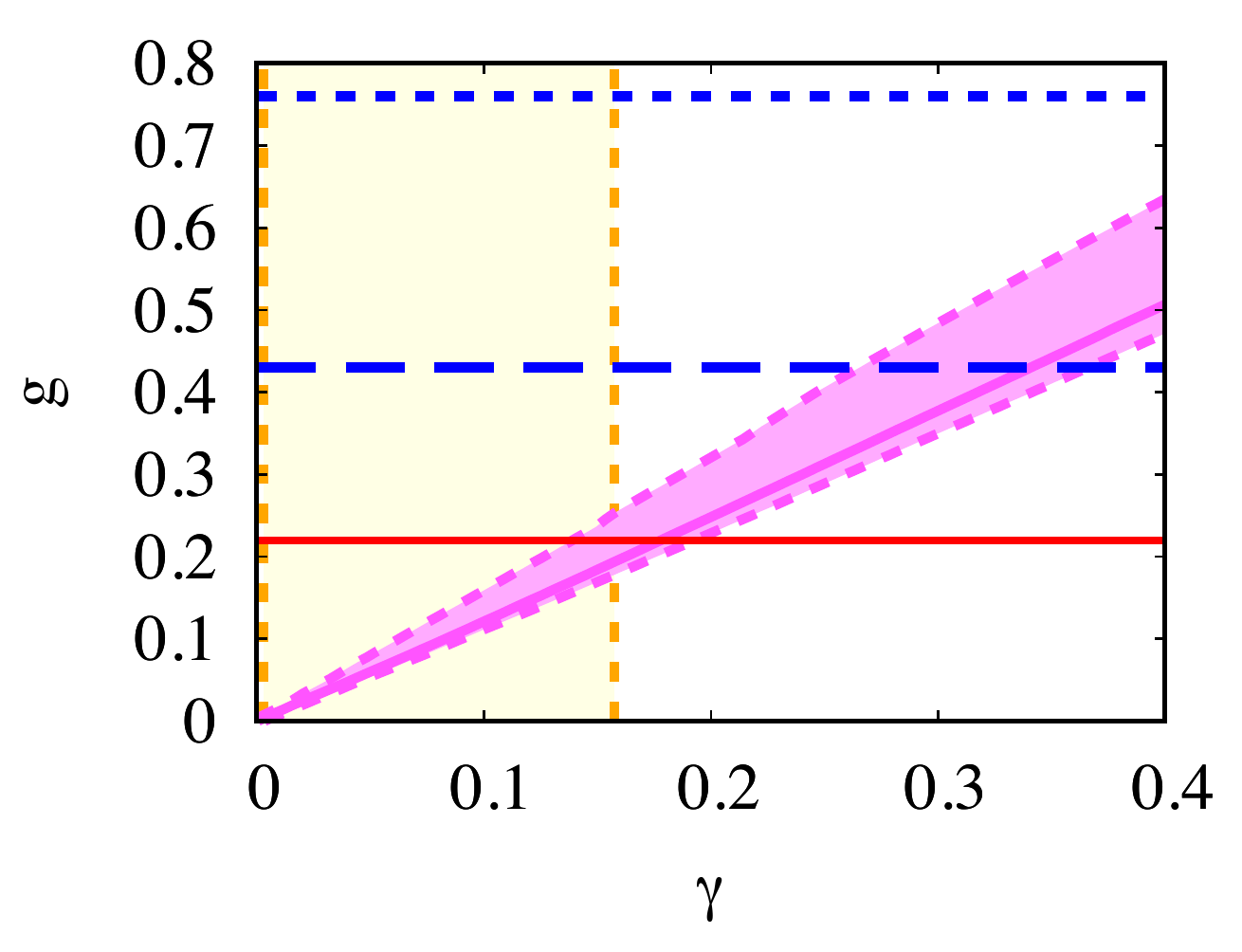}
\caption{Growth function $g$ as a function of the slope of the normalized local main sequence $\gamma$, adopting a two-component (bulge+disc) model for the stellar structure of SFGs. The solid line (shaded area) in magenta indicates the median (16th-84th percentile range) of $g$ for the observed distribution of structural parameters from the S$^4$G survey. The observed range of $\gamma$ (Section \ref{subsec:implications}) is indicated as a yellow shaded band demarcated by dashed vertical lines. Horizontal lines mark for reference the predictions of three relevant models, with line-styles as in Figure \ref{fig:mainresult}. Observational constraints (at the intersection of the two shaded areas) favour a close to critical growth function (solid red line), in good agreement with the single-Sersic description in Figure \ref{fig:mainresult}.}\label{fig:2comp}
\end{figure}

The check is performed on S$^4$G late-type galaxies (defined as in Section \ref{subsec:implications}), further selected to have maximum reliability (quality flag equal to 5) two component (bulge+disc) decompositions as reported by \cite{Salo+15}. For each galaxy, the growth function $g$ is computed numerically, using equations \eqref{g_sSFR} and \eqref{normalisedmainsequence}, as a function of the slope of the normalized local main sequence $\gamma$. The results are shown in Figure \ref{fig:2comp}, where the solid line and the shaded area in magenta indicate, respectively, the median and the 16th-84th percentile range of the growth function across the sample.

As expected, the growth function vanishes ($g=0$) for a perfectly linear local main sequence ($\gamma=0$), regardless of the galaxy structure (thence the vanishing scatter in the origin) and increases (together with its scatter) with increasing $\gamma$.

Horizontal lines in Figure \ref{fig:2comp} indicate, for reference, the theoretical growth function for the same three models (and following the same line-styles) as in Figure \ref{fig:mainresult}, while the vertical shaded band in yellow highlights the range of $\gamma$ favoured by observations of the local main sequence, as described in Section \ref{subsec:implications}.

Among the three models shown and for local main sequence slopes within the indicated region, the growth function of S$^4$G late-type galaxies is consistent with being close to critical (red solid line) and formally inconsistent with the other two models, in good agreement with the results of Section \ref{subsec:implications}.

The exercise confirms that a single-parameter description of stellar structure, based on the global Sersic index $n$, is sufficient for the purposes of this work. At the same time, it exemplifies how the formalism introduced in Section \ref{sec:RadialGrowthRate} for the calculation of the radial growth rate (and thence the growth function) can be easily applied to more elaborate multi-component descriptions, if needed.

\bsp	
\label{lastpage}
\end{document}